\documentclass[prd,twocolumn,nofootinbib,notitlepage,raggedbottom]{revtex4-1}

\usepackage{graphicx}
\usepackage{dcolumn}
\usepackage{bm}
\usepackage{color}
\usepackage{hyperref}
\usepackage{amsmath, mathtools}
\usepackage{amssymb}
\usepackage{wasysym}
\usepackage{graphicx} 
\usepackage{booktabs} 
\usepackage{pdfsync}
\usepackage{verbatim}
\usepackage[utf8]{inputenc}
\usepackage{latexsym}
\usepackage{dsfont}

\def\sec#1{\section{#1} }
\def\ssec#1{\subsection{#1} }

\def\({\left(}
\def\){\right)}
\def\[{\left[}
\def\]{\right]}

\def\a{\alpha}

\def\f#1#2{\frac{#1}{#2}}

\def\Ga{\Gamma}

\def\l{\lambda}
\def\L{\Lambda}

\def\p{\pi}
\def\r{\rho}

\def\s{\sigma}

\def\th{\theta}

\def\cmsg{\text{~cm}^2/\text{g}}
\def\gcmc{\text{g/cm}^3}

\def\MeV{\text{MeV}}

\def\cm{\text{cm}}

\def\<{\langle}
\def\>{\rangle}

\def\sX{\s_\text{\tiny X}}

\def\MX{M_\text{\tiny X}}

\def\RX{R_\text{\tiny X}}

\def\rX{\r_\text{\tiny X}}

\def\sm{\sX/\MX}

\begin{document}

\author{David M. Jacobs}
\email{Corresponding author: dmj15@case.edu}
\affiliation{Astrophysics, Cosmology and Gravity Centre,\\
Department of Mathematics and Applied Mathematics,\\
University of Cape Town\\
Rondebosch 7701, Cape Town, South Africa}

\author{Gwyneth Allwright}
\affiliation{Astrophysics, Cosmology and Gravity Centre,\\
Department of Mathematics and Applied Mathematics,\\
University of Cape Town\\
Rondebosch 7701, Cape Town, South Africa}

\author{Mpho Mafune}
\affiliation{Astrophysics, Cosmology and Gravity Centre,\\
Department of Mathematics and Applied Mathematics,\\
University of Cape Town\\
Rondebosch 7701, Cape Town, South Africa}

\author{Samyukta Manikumar}
\affiliation{Astrophysics, Cosmology and Gravity Centre,\\
Department of Mathematics and Applied Mathematics,\\
University of Cape Town\\
Rondebosch 7701, Cape Town, South Africa}

\author{Amanda Weltman}
\affiliation{Astrophysics, Cosmology and Gravity Centre,\\
Department of Mathematics and Applied Mathematics,\\
University of Cape Town\\
Rondebosch 7701, Cape Town, South Africa}

\title{Primordial $^4\text{He}$ constraints on inelastic macro dark matter revisited}

\begin{abstract}
At present, the best model for the evolution of the cosmos requires that dark matter make up approximately $25\%$ of the energy content of the Universe. Most approaches to explain the microscopic nature of dark matter, to date, have assumed its composition to be of intrinsically weakly interacting particles; however, this need not be the case to have consistency with all extant observations. Given decades of inconclusive evidence to support any dark matter candidate, there is strong motivation to consider alternatives to the standard particle scenario. One such example is macro dark matter, a class of candidates (macros) that could interact strongly with the particles of the Standard Model, have large masses and physical sizes, and yet behave as dark matter. Macros that scatter completely inelastically could have altered the primordial production of the elements, and macro charge-dependent constraints have been obtained previously. Here we reconsider the phenomenology of inelastically interacting macros on the abundance of primordially produced $^4\text{He}$ and revise previous constraints by also taking into account improved measurements of the primordial $^4\text{He}$ abundance. The constraints derived here are limited in applicability to only leptophobic macros that have a surface potential $V(\RX)\gtrsim 0.5\, \MeV$. However, an important conclusion from our analysis is that even neutral macros would likely affect the abundance of the light elements. Therefore, constraints on that scenario are possible and are currently an open question.
\end{abstract}


\maketitle
\flushbottom

\sec{Introduction}

We are presently in a dark age of modern cosmology, not because of a lack of agreement between the $\L\text{CDM}$ model of the Universe and measurements of the cosmos, but because of the lack of understanding of the basic constituents in the model. The unknown ``dark" components of the model, namely, dark energy and dark matter, apparently account for upwards of $95\%$ of the energy content of the Universe. Despite difficulties in attributing its value to quantum field-theoretic contributions, dark energy has so far been successfully described in terms of a cosmological constant that pervades all of spacetime. Dark matter, on the other hand, would appear to be composed of some type of material substance whose theoretical description is essentially unknown.

Meanwhile, the Standard Model of particle physics has been extremely successful at describing nearly all (nongravitational) fundamental processes observed in nature so far. Although dark matter cannot be made of individual particles in the Standard Model--their interaction strengths are simply too strong (see, e.g., \cite{Boehm:2000gq, Wilkinson:2013kia})--composite objects with a Standard Model origin can evade extant bounds.  Such models have been proposed before, most notably by Witten \cite{Witten:1984rs} and in several subsequent works (e.g., \cite{DeRujula:1984ig, Farhi:1984qu, Lynn:1989xb, Zhitnitsky:2002nr, Zhitnitsky:2002qa, Zhitnitsky:2006vt, Lynn:2010uh, Labun:2011wn}), wherein the dark matter candidates would have nuclear density and may be (nearly) a result of entirely Standard Model physics. Other similarly massive candidates include primordial black holes \cite{Carr:1974nx}, and other more exotic candidates which may be found in the literature (e.g., \cite{Kusenko:1997si, Khlopov:2013ava, Murayama:2009nj, Derevianko:2013oaa, Stadnik:2014cea}).

Whether or not they have a Standard Model origin, \emph{macros} \cite{Jacobs:2014yca} remain a viable dark matter candidate and an alternative to the comparatively light candidates, such as weakly interacting massive particles or axions (see, e.g., \cite{Peter:2012rz} for a brief review of those candidates). Macros represent the general class of macroscopically large dark matter candidates whose interaction strengths are given by their geometric cross sections.   Roughly speaking, their viability depends on the following criteria: they are massive enough that their expected signal rate is below the relevant experimental sampling rate, e.g., months, decades, millennia, etc.; their reduced cross section, $\sm$, is small enough to evade astrophysical and cosmological bounds; they are not so massive to have already been detected through gravitational lensing. At present, this has still left a rather large window of possibilities between about $50$ g to $10^{17}$ g, and another small window between $10^{20}$ g and roughly $4\times 10^{24}$ g \cite{Jacobs:2014yca, Jacobs:2015csa}.

Macros are modeled to interact completely elastically or inelastically, and the phenomenology of each case can differ greatly. In the latter case, baryons would be absorbed by macros, thereby affecting the primordial production of the light elements, for example. This was first considered in \cite{Jacobs:2014yca}, wherein the effects on the $^4\text{He}$ production was calculated and compared to abundance measurements to obtain a constraint on $\sm$. It was determined in that analysis that for nuclear-inspired models, if the macro radius, $\RX\gtrsim 10^{-10}\cm$ then the surface potential, $V(\RX)\simeq0$; this is because the surrounding plasma effectively neutralizes the macro charge.  Additionally, it may perhaps be more natural for an underlying theory to predict neutral macroscopic dark matter.  However, it was concluded that they would have little impact on the production of the light elements because protons and neutrons would be absorbed at approximately equal rates; the analysis in \cite{Jacobs:2014yca} therefore only focused on macros that would be leptophobic, or at least did not admit electrons or positrons that would effectively neutralize the macros. We know of no such models at present; however, this is possible in principle.


In this work we continue to focus on macros with radius, $\RX\gg 10^{-10}\cm$ that are leptophobic, and make three improvements to the earlier analysis: we account for possible $^4\text{He}$ absorption \emph{after} its initial production; updated $^4\text{He}$ abundance determinations given in \cite{Aver:2015iza} are used to infer an improved constraint on $\sm$ in the context of inelastically interacting macro dark matter;  we discuss the limitations of our results due to additional physical effects that were not previously considered, primarily Debye screening of the primordial plasma. The major insight gained from this analysis, however, is that a more complete analysis of big bang nucleosynthesis (BBN) may well provide a constraint for the most important and interesting case of neutral, i.e. $V(\RX)=0$, macro dark matter. 



\sec{Theory}

\ssec{Limitations}

First, macros could have interfered with the weak interaction processes that determined the temperature at onset of BBN, and hence the neutron-to-proton ratio. To avoid this complication, for example, the rate at which neutrons were absorbed by macros must have been much smaller than the rate for the process $n + \nu_e\to p+e^-$, i.e. $\Gamma_{nX}\ll \Gamma_{n\nu}$. The expression for $\Gamma_{nX}$ is given below in equation \eqref{Macro_Rate}, while $\Gamma_{n\nu}$  can be found in a standard text that includes primordial nucleosynthesis (e.g., \cite{Mukhanov:2005sc}). The inequality is approximately 
\begin{equation}
5 \times 10^4 \times T_9^{7/2} \(\f{\sX}{\MX} \f{\text{g}}{\text{cm}^2}\) \ll  3\times10^{-6}\, T_9^5\,,
\end{equation}
or
\begin{equation}
\(\f{\sX}{\MX} \f{\text{g}}{\text{cm}^2}\) \ll  6\times10^{-11}\, T_9^{3/2}\,,
\end{equation}
and comparison with the rates of other weak processes leads to similar results. In the standard scenario weak freeze-out would have occurred at approximately $T_F\simeq10$, so we will need to obtain a constraint of $\sm\ll10^{-9}\cmsg$  to justify our assumption that freeze-out is standard.

Second, there are two important length scales that were neglected in the earlier analysis associated with the finite plasma density, and this has significant implications for the electromagnetic enhancement of the cross section. The free-streaming length of the nuclides, $\l_\text{free}$, is not infinite; it is shortened due to collisions with ambient particles, the dominant of which are photons. Because each collision is a random walk in momentum space, it takes $N_c\sim \(q/T\)^2$ collisions for a proton of momentum, $q$, with photons of momentum, $T$, to be displaced in momentum space by ${\cal O}(q)$. From thermodynamic considerations, $N_c\sim m_p/T$.

The collision time for a proton with an individual photon is approximately $\a^{-2} T^{-1}$, which follows from a cross section of approximately $\a^2/T^2$ and a photon number density of roughly $T^3$. The free-streaming length is the product of the collision time, thermal velocity ($\sqrt{T/m_p}$), and number of collisions per significant momentum change; therefore, 
\begin{equation}
\l_\text{free}\sim \f{1}{\a^2 T} \(\f{m_p}{T}\)^{1/2}\,.
\end{equation}

The Debye length, $\l_\text{D}$, is the (exponential) decay length of the electric potential due to the screening effect of the plasma,
\begin{equation}
\l_\text{D}=\(\f{4\p \a \bar{n}}{T} \)^{-1/2}\,,
\end{equation}
where $\bar{n}=\sum_i n_i$ is the sum of the number densities of the charged species. For $T\gtrsim m_e$ the plasma charge density is dominated by electrons and positrons, while afterward it is dominated by protons and electrons; therefore,
\begin{equation}
\bar{n}\sim
\begin{cases}
0.4\,T^3~~~~~~~~~&{(T\gtrsim m_e)}\\
2\times10^{-10}\,T^3 &{(T\lesssim m_e)}\,,
\end{cases}
\end{equation}
which means
\begin{equation}
\l_\text{D}\sim T^{-1}
\begin{cases}
5~~~~~~~~~&{(T\gtrsim m_e)}\\
2\times10^{5} &{(T\lesssim m_e)}\,.
\end{cases}
\end{equation}
or
\begin{equation}
\l_\text{D}\sim T_9^{-1}
\begin{cases}
10^{-9} \text{cm} ~~~~~&{(T\gtrsim m_e)}\\
4\times10^{-5} \text{cm}  &{(T\lesssim m_e)}\,.
\end{cases}
\end{equation}
One may check that $\l_\text{D}\ll\l_\text{free}$ during BBN; we, therefore, only concern ourselves with $\l_\text{D}$. If we restrict our attention to $\RX\gtrsim 10^{-5}\cm$, then the potential is always Debye screened. The constraints obtained then also apply for the case $\RX\lesssim 10^{-5}\cm$, as they would be conservative.

\ssec{Primordial $^4\text{He}$ simplified}
The relative abundance of protons and neutrons in the early Universe, we have explicitly assumed, is governed by the weak interaction until the ambient temperature drops below the freeze-out temperature, $T_F\simeq 1\,\MeV$, when the relative neutron-to-proton abundance is determined by the ratio of their masses and the temperature $T_F$. Protons and neutrons can interact and form deuterium; however, it cannot be produced in significant amounts until after the number of ambient photons with energies greater than the deuterium binding energy drops low enough. After this so-called deuterium bottleneck, deuterium and heavier elements can start being produced in significant quantities once the bottleneck breaks at a temperature, $T_B$.

The production of $^4\text{He}$ is very efficient after this point, and it is a reasonable approximation to count all remaining neutrons at $T_B$ as having ended up in $^4\text{He}$ (see, e.g., \cite{Mukhanov:2005sc}). Hence, the primordial $^4\text{He}$ mass fraction, $X_4$, is approximately
\begin{align}
X_4 \simeq \frac{2n_n}{n_n + n_p}\,,
\end{align}
where $n_p$ and $n_n$ are the proton and neutron abundances, respectively.

However, in the presence of macros, some protons and neutrons may not participate in the primordial element production if, for example, they are absorbed or they are catalyzed to decay (see, e.g., \cite{Madsen:1985zx}). If macros are charged, then they would have absorbed protons at a different rate than neutrons, and thus the value of $X_4$ at the temperature $T_B$ would differ from the canonical value.

Additionally, an important point that we wish to clarify here is that $^4\text{He}$ and protons would be absorbed after the breaking of the deuterium bottleneck, at temperatures lower than $T_B$.  During this later stage, one may write
\begin{equation}
X_4 = \frac{4n_{4}}{4n_{4} + n_{p,f}}\,,
\end{equation}
where $n_4$ is the  $^4\text{He}$ number density and $n_{p,f}$ is the number density of free protons, i.e. those not contained in $^4\text{He}$. We, therefore, perform an analysis that is consistent for all relevant temperatures by defining
\begin{equation}
X_4 = \frac{2}{1 + \alpha}\,,
\end{equation}
where
\begin{equation}
\a\equiv
\begin{cases}
\f{{\cal N}_p}{{\cal N}_n}~~~~~~&(T_B\leq T\leq T_F)\\
1 + \f{1}{2}\f{{\cal N}_{p,f}}{{\cal N}_4} &(T\leq T_B)\,,
\end{cases}
\end{equation}
where the ${\cal N}_i\equiv a(t)^3 n_i$ are comoving number densities.

What follows below is a straightforward generalization of the analysis presented in \cite{Jacobs:2014yca}. Accounting for neutron decay and absorption by macros, and ignoring all composite elements besides $^4\text{He}$, the evolution of the relevant species is determined by the following equations before the deuterium bottleneck break,
\begin{align}
\dot{{\cal N}_n} &= -\(\Gamma_n + \Gamma_{nX} \){\cal N}_n\\
\dot{{\cal N}_p} &=  +\Gamma_n {\cal N}_n  - \Gamma_{pX}{\cal N}_p\,,
\end{align}
and after the bottleneck break,
\begin{align}
\dot{\cal N}_4&=  - \Gamma_{4X}{\cal N}_4\\
\dot{\cal N}_{p,f}&=- \Gamma_{pX}{\cal N}_{p,f}\,.
\end{align}
Here, we take $\Gamma_n\simeq (880.3\,\text{s})^{-1}$ as the neutron decay rate \cite{Agashe:2014kda}, and the absorption rates of all species can be written in terms of the neutron absorption rate, as described below. Since a neutron is neutral, its macro absorption rate is simply given by
\begin{align}\label{Macro_Rate}
\Gamma_{nX}&=\<\f{\rX}{\MX} \sX v\>\notag\\
&=5.1 \times 10^4 \times T_9^{7/2} \f{\sX}{\MX} \f{\text{g}}{\text{cm}^2 \text{ s}}\,,
\end{align}
where we have used the thermally averaged neutron velocity $v_n=\sqrt{8T/(\pi m_n)}$ and inserted $\rX =3H_0^2/(8 \pi G)\Omega_c\(T/T_0\)^3=0.93\times 10^{-3} \Omega_c h^2 T_9^3 ~\gcmc$ with the Planck value of $\Omega_c h^2=0.1188$ \cite{Ade:2015xua}.

Charged particles require a bit more care; consider an $i$-type particle with mass, velocity, and charge, $m_i, v_i$, and $q_i$, respectively, where $q_i$ is given in units of proton charge, $+e$.  Because we limit ourselves to the case $\RX\gtrsim\l_\text{D}$ wherein the Coulomb effects are screened, the effective cross section can be considered,
\begin{equation}
\s_\text{\tiny X,eff}=\sX\,\Theta(E_i - q_i V(\RX) )\,,
\end{equation}
where $\Theta(x)$ is the Heaviside theta function. This must be thermally averaged along with the velocity, resulting in
\begin{equation}
\<\s_\text{\tiny X,eff}v_i\>=\sX\<v_i\>
\end{equation}
for $q_i V(\RX)<0$, and
\begin{equation}
\<\s_\text{\tiny X,eff}v_i\>=\sX\<v_i\>\times
e^{-\f{q_iV(\RX)}{T}}\(1+\f{q_iV(\RX)}{T}\)\,
\end{equation}
for $q_i V(\RX)>0$.
It then follows that the absorption rate for this type of particle may be written in terms of the neutron absorption rate as
\begin{equation}
\Gamma_{iX}=\sqrt{\f{m_n}{m_i}}\Gamma_{nX}\,,
\end{equation}
for $q_i V(\RX)<0$, while we find\footnote{We have also checked the effect of quantum tunneling. The exponent of the quantum process is approximately $\l_\text{D} \sqrt{2m q_i V(\RX)}$, which is much greater than for the thermal process, namely $q_i V(\RX)/T$; therefore tunneling is negligible compared to the classical (thermal) barrier hopping.}
\begin{equation}
\Gamma_{iX}=\sqrt{\f{m_n}{m_i}}\Gamma_{nX} 
 e^{-q_iV(\RX)/T}\(1+\f{q_iV(\RX)}{T}\)
\end{equation}
for $q_i V(\RX)>0$. Not only is the absorption rate charge dependent, but because the thermal velocities are inversely proportional to $\sqrt{m_i}$, the absorption rates of different elements differ by factors of ${\cal O}(1)$, even if the macro surface potential is zero.  As we discuss below, this is an important correction to the original analysis in \cite{Jacobs:2014yca} in that even macros with a neutral core could affect the abundance of $^4\text{He}$.

Given the discussion above, $\a(t)$ obeys two different evolution equations, depending on the era:
\begin{equation}
\dot{\a}(t)=
\begin{cases}
 \Ga_n + \(\Ga_n +\Ga_X\)\a(t),~~~~&(T_F\geq T\geq T_B)\\
\( \a(t)-1\)\(\Gamma_{4X}-\Gamma_{pX}\), &(T_B\geq T)\,.
\end{cases}
\end{equation}
In the first line, we assume the initial condition for $\a$ at $T_F$ to be set entirely by standard weak-interaction physics; this evolution ends at $T_B$ and sets the initial condition for the evolution specified in the second line.

Assuming macros to have a perturbatively small effect on $X_4$, we find the deviations to the standard value
\begin{align}\label{X_4_Macro_effect}
\Delta X_4^\text{macro}&\equiv X_4-X_4^\text{std}\notag\\
&\simeq -\f{2}{\(1+\a^\text{std}\)^2}\[\a^\text{std}(a+c)-(b+c)\]\,,
\end{align}
where we will use the standard value $\a^\text{std}\simeq7$  and
\begin{align}
a&=\int_{t_F}^{t_B}dt~\(\Ga_{nX}-\Ga_{pX}\)\\
b&=  e^{\int^{t_B}_{t_F}dt\Ga_n} \notag\\
&~~~\times\int_{t_F}^{t_B}d\tilde{t}~\Ga_n~e^{-\int_{t_F}^{\tilde{t}}dt~\Ga_n} \int_{t_F}^{\tilde{t}}dt~\(\Ga_{nX}-\Ga_{pX}\)\\
c&=\int_{t_B}^{\infty}dt~\(\Ga_{4X}-\Ga_{pX}\)\,.
\end{align}

To perform these integrals, we change the integration variable to temperature using the time-temperature relation
\begin{equation}
t=\f{\th}{T_9^2} ~\text{s}\,,
\end{equation}
where $T_9$ is the temperature defined in units of $10^9$K and $\theta$ depends on the number of relativistic degrees of freedom. Following \cite{Esmailzadeh:1990hf}, we find
\begin{equation}
\th=
\begin{dcases}
&99.4,~~~~ T_9>5\\
&178, ~~~~~T_9<1\,,
\end{dcases}
\end{equation}
assuming the standard value of $N_\text{eff}=3.046$. In what follows, we use the values $T_{9,F}=9.1$ and $T_{9,B}=1$ as in \cite{Esmailzadeh:1990hf} and numerically determine $a$, $b$, and $c$ for different values of the macro surface potential, $V(\RX)$ \footnote{We have checked that variations of $T_B$ by 10\% do not affect the values of our asymptotic constraints by more than 5\%.}.
Integrating through this range requires an interpolation of $\th(T_9)$ in the region $1\leq T_9 \leq 5$; to do this we follow \cite{Jacobs:2014yca} by choosing a hyperbolic tangent centered around $T_9=2$,
\begin{equation}\label{theta_interpolation}
\th(T_9) \simeq \th_\text{max} - \f{1}{2}\(\th_\text{max} - \th_\text{min} \)\(\tanh{\[T_9-2\]} +1\), 
\end{equation}
where $\th_\text{max}=178,\, \th_\text{min}=99.4$. Equation \eqref{theta_interpolation} is sufficient to match all of the values quoted in Table 15.5 of Reference  \cite{Weinberg} to an error of less than 10\%. Since the integrals are performed over a range of temperatures for which the interpolation is much more accurate than 10\%, we expect the error from this interpolation to result in errors that are well below 10\% on the constraints derived here.\footnote{In \cite{Weinberg} there were only two neutrino species used to calculate Table 15.5 therein; to compare, $\th_\text{max}$ and $\th_\text{min}$ must be corrected to account for this smaller number of relativistic fermions.}

Last, we note that there are other physical effects that could alter the measured value of $X_4$, such as the existence
of extra neutrino species or errors in the measurement of the baryon fraction, $\Omega_b$. Although these would be degenerate with the macro effects that we consider, we find that they are negligible given the macro constraints that we obtain here.

\sec{Results}

The primary sources of data for inferring primordial abundances of $^4\text{He}$ are low metallicity, extragalactic H II regions--large low density clouds containing significant amounts of ionized hydrogen and helium. Information about the hydrogen and helium abundance is gathered from studying their emission spectra \cite{Aver:2015iza}.

\begin{figure}[h]
  \begin{center}
    \includegraphics[scale=.45]{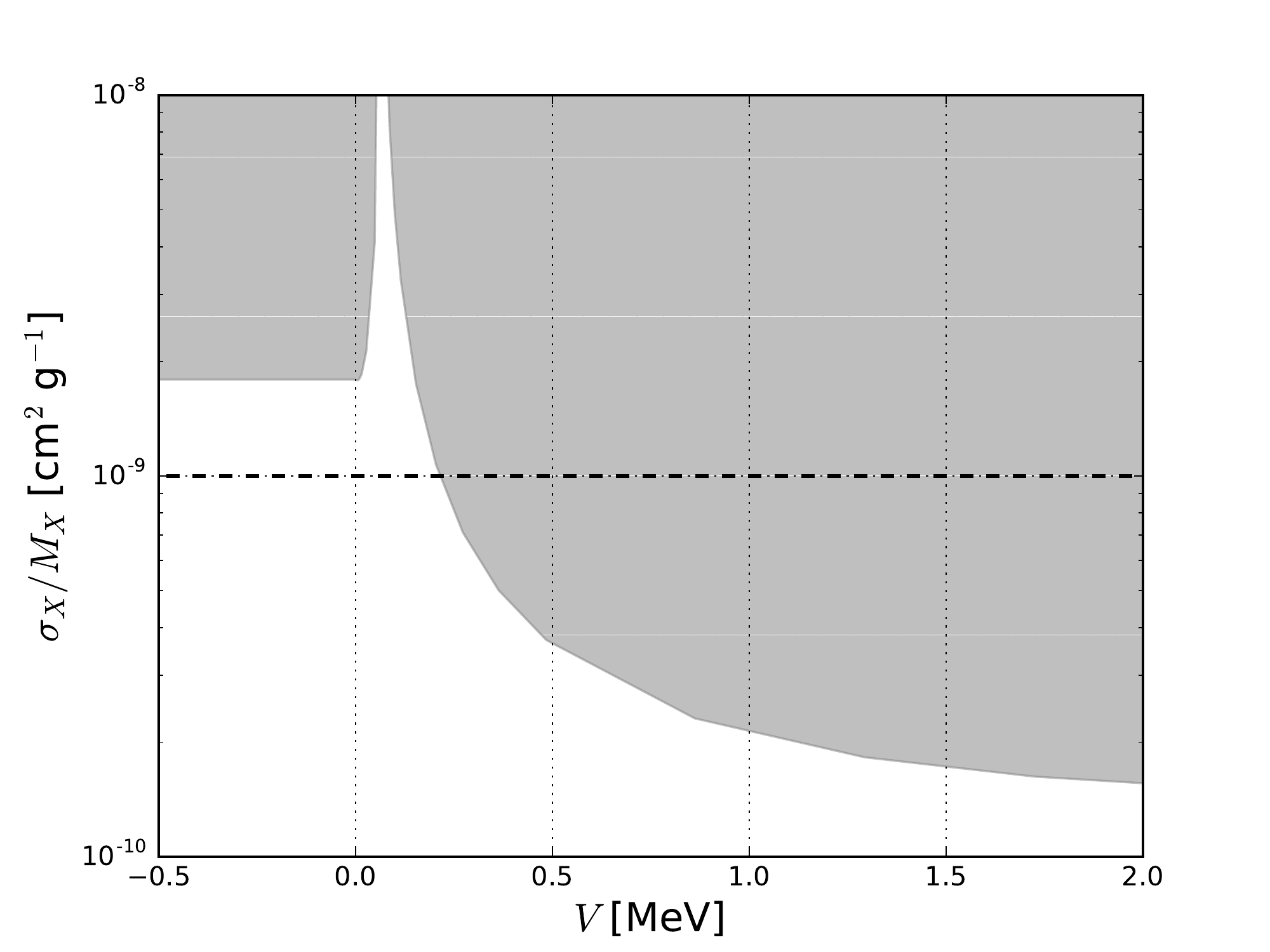}
  \end{center}
\caption{New constraints on $\sm$ based on primordial $^4\text{He}$.  The horizontal dashed line indicates the value of $\sm=10^{-9} \cmsg$ above which the weak interaction freeze-out would have likely been altered from the standard scenario. Therefore, at present, we do not trust the constraint above this line.}
\label{fig2}
\end{figure}

However, the presently observed $^4\text{He}$ abundance differs from the abundance in the early Universe because of stellar processing. To account for this, it is important to quantify the dependence of $X_4$ on the metallicity (the O/H abundance ratio) of the region in which it is observed.  The primordial $^4\text{He}$ abundance is then inferred by extrapolating the measurements to zero metallicity \cite{Izotov:2003xn}. Recently, using data presented in \cite{Izotov:2014fga}, Aver \emph{et al.} \cite{Aver:2015iza} have made improved $^4\text{He}$ abundance measurements of individual galaxies by including the He I $\l$10830 emission line in their analyses. As a result, they have determined the primordial $^4\text{He}$ abundance to be
\begin{equation}
X_4^\text{obs}=0.2449 \pm 0.0040\,,
\end{equation}
reducing the error by greater than 50\% compared to previous determinations.  At the same time, the Standard Model theoretical prediction \cite{Cyburt:2015mya} is
\begin{equation}
X_4^\text{theory}=0.2470 \pm 0.0002\,.
\end{equation}
Given this small theoretical uncertainty, one can conclude that any deviations from the canonical prediction caused by macros is constrained to be
\begin{equation}
-0.006\leq  \Delta X_4^\text{macro} \leq 0.002\,.
\end{equation}
This observational bound, along with numerical solutions to \eqref{X_4_Macro_effect}, have enabled us to put a constraint on the ratio $\sm$ as a function of $V(\RX)$, illustrated in Fig. \ref{fig2}. In the figure, we indicate the regime where our constraints become less robust with a dashed line indicating where $\sm=10^{-9} \cmsg$; for the weaker constraints found above that line it is possible that the weak interaction freeze-out could have been significantly affected, e.g., the $n/p$ ratio altered, and hence this analysis is not reliable because our assumptions are no longer valid. 
For $V(\RX)\lesssim0$, we would find the constraint is approximately
\begin{equation}\label{neg_and_zero_V_limit}
\f{\sX}{\MX} \lesssim 1.8 \times 10^{-9} \cmsg \,.
\end{equation}
Unfortunately, this bound is not strong enough to be consistent with our starting assumptions, as mentioned above. For $V(\RX)\gtrsim 1\,\MeV$ we find the constraint to approach
\begin{equation}\label{large_V_limit}
\f{\sX}{\MX} \lesssim 1.4 \times 10^{-10} \cmsg\,,
\end{equation}
an improvement of approximately a factor of 2 over previous results obtained in \cite{Jacobs:2014yca}. These results are qualitatively similar to that work; however, upon close inspection of Fig. \ref{fig2}, one can see that in the vicinity of $V(\RX)\simeq 0.06\,\MeV$ the bounds become weak and essentially vanish.\footnote{Changes to the value of $T_B$ by $\pm10\%$ would only affect this critical value by approximately $0.001\,\MeV$.} The reason for this is as follows. For $V(\RX)=0$, the initial production of $^4\text{He}$ is approximately unaffected by macros because neutrons would have been absorbed at nearly the same rate as protons. However, after its initial production, $^4\text{He}$ would have been absorbed at only half the rate of protons, so abundance measurements at the present time would be bigger in the presence of macros than in the standard prediction. On the other hand, for a particular positive value of $V(\RX)$ at which the initial value of $X_4$ is smaller than the standard prediction\footnote{This would happen because of an excess of neutrons absorbed.}, $X_4$ would then increase back to the standard value during its later evolution because of the effect mentioned above. In this case, our macro bounds would therefore vanish.

Another salient feature of Fig. \ref{fig2} is that there would have been a finite constraint on $\sm$ for $V(\RX)=0$ given by \eqref{neg_and_zero_V_limit} if the constraint could be trusted. While we cannot fully trust the constraint there, it is nevertheless encouraging because a more sophisticated analysis may well provide a tighter bound on this most interesting case; a constraint that is even an order of magnitude lower would be sufficient to be reliable. However, this is beyond the scope of this work.



\sec{Discussion}

Here we have revisited the possibility of using the theory of BBN and the measured primordial abundance of $^4\text{He}$ to constrain the properties of inelastically interacting macro dark matter. Our results are constraints on the reduced cross section, $\sm$ as a function of the macro surface potential, $V(\RX)$. Improvements over previous results were possible for three reasons: an improved study of the absorption rates of different nuclides, reduced uncertainties on the observed primordial $^4\text{He}$ abundance, and a more thorough analysis of the physical effects relevant during BBN, especially Debye screening. Our constraint of $\sm \lesssim 1.4 \times 10^{-10} \cmsg$ applies for leptophobic models wherein $V(\RX)\gtrsim 0.5\,\MeV$.

In addition to these improved asymptotic bounds on $\sm$ compared to \cite{Jacobs:2014yca}, we have found that, because elements of different masses are absorbed at significantly different rates, nontrivial constraints can likely be obtained for $V(\RX)=0$; this more interesting (and probably more likely) possibility does not require macros to be leptophobic. 
In order to provide a robust constraint for $V(\RX)\simeq 0$, an improved constraint on $\sm$ must be obtained. It appears that a consideration of the effect on deuterium, and possibly several of the other light elements, should be taken into account. Although a full BBN analysis is beyond the scope of the work presented here, our results are significant because they suggest that nontrivial constraints might be obtainable for all inelastically interacting macros, regardless of their surface potential. This is the subject of a work in progress \cite{Jacobs:IP}.  

\section*{Acknowledgements}
The authors would like to sincerely thank Glenn Starkman for bringing the issues of free-streaming and Debye screening to their attention. They also thank Yin-Zhe Ma for discussions. One of the authors (D.M.J.) would like to acknowledge support from the Claude Leon Foundation. This work is based on the research supported by the South African Research Chairs Initiative of the Department of Science and Technology and National Research Foundation of South Africa as well as the Competitive Programme for Rated Researchers (Grants No. 91552 and No. 81134) (A.W., G.A., and M.M.).  Any opinion, finding, conclusion, or recommendation expressed in this material is that of the authors and the NRF does not accept any liability in this regard.

\bibliographystyle{apsrev4-1}
\bibliography{Macro_DM_bib}

\end{document}